\documentclass[aps,prl,citeautoscript,showpacs,reprint,amsmath,amssymb]{revtex4-1}

\usepackage{graphicx}
\usepackage{mathtools}
\usepackage{times}
\usepackage{bm}
\usepackage{ulem}
\usepackage[
            colorlinks,linkcolor=blue,
            citecolor=blue,
            pdfstartview=FitH,
            plainpages=false]
            {hyperref}
                        
\usepackage{color}
\usepackage{soul}

\begin{document}

\title{A magnetic Weyl semimetallic phase in thin films of Eu$_2$Ir$_2$O$_7$}

\author{Xiaoran Liu$^{1,2}$}
\email{xiaoran.liu@rutgers.edu}
\author{Shiang Fang$^{1,3}$}
\email{shiangfang913@gmail.com}
\author{Yixing Fu$^{1,3}$}
\author{Wenbo Ge$^1$}
\author{Mikhail Kareev$^1$}
\author{Jong-Woo Kim$^4$}
\author{Yongseong Choi$^4$}
\author{Evguenia Karapetrova$^4$}
\author{Qinghua Zhang$^2$}
\author{Lin Gu$^2$}
\author{Eun-Sang Choi$^5$}
\author{Fangdi Wen$^1$}
\author{Justin H.  Wilson$^{1,3}$}
\author{Gilberto Fabbris$^4$}
\author{Philip J. Ryan$^4$}
\author{John Freeland$^4$}
\author{Daniel Haskel$^4$}
\author{Weida Wu$^1$}
\author{J. H. Pixley$^{1,3,6,7}$}
\author{Jak Chakhalian$^1$}

\affiliation{$^1$ Department of Physics and Astronomy, Rutgers University, Piscataway, New Jersey 08854, USA}
\affiliation{$^2$ Beijing National Laboratory for Condensed Matter Physics and Institute of Physics, Chinese Academy of Science, Beijing 10019, P. R. China.}
\affiliation{$^3$ Center for Materials Theory, Rutgers University, Piscataway, New Jersey 08854, USA.}
\affiliation{$^4$ X-ray Science Division, Argonne National Laboratory, Argonne, Illinois 60439, USA.}
\affiliation{$^5$ National High Magnetic Field Laboratory, Florida State University, Tallahassee, Florida 32310, USA.}
\affiliation{$^6$ Center for Computational Quantum Physics, Flatiron Institute, 162 5th Avenue, New York, NY 10010.}
\affiliation{$^7$ Physics Department, Princeton University, Princeton, New Jersey 08544, USA.}

\date{\today}

\begin{abstract}
The interplay between electronic interactions and strong spin-orbit coupling is expected to create a plethora of fascinating correlated topological states of quantum matter. Of particular interest are magnetic Weyl semimetals originally proposed in the pyrochlore iridates, which are only expected to reveal their topological nature in thin film form. To date, however, direct experimental demonstrations of these exotic phases remain elusive, due to the lack of usable single crystals and the insufficient quality of available films. Here, we report on the discovery of the long-sought magnetic Weyl semi-metallic phase in (111)-oriented Eu$_2$Ir$_2$O$_7$ high-quality epitaxial thin films. The topological magnetic state shows an intrinsic anomalous Hall effect with colossal coercivity but vanishing net magnetization, which emerges below the onset of a peculiar magnetic phase with all-in-all-out antiferromagnetic ordering. The observed anomalous Hall conductivity arises from the non-zero Berry curvature emanated by Weyl node pairs near the Fermi level that act as sources and sinks of Berry flux, activated by broken cubic crystal symmetry at the top and bottom terminations of the thin film.   
\end{abstract}

\maketitle

Quantum materials with nontrivial band topology have rapidly developed as a central theme in condensed matter physics \cite{Balents_ARCMP_2014, Balents_NatPhys_2010, Sato_RPP_2017,Tokura_NP_2017}. The focus has recently shifted towards the search and discovery of gapless topological compounds, whose low-energy band structures can harbor Weyl or Dirac fermionic excitations in three-dimensions (3D)~\cite{Wan_PRB_2011,Nagaosa_PRL_2014,Fiete_PRL_2017,Kim_SciRep_2016,Yan_ARCMP_2017,Schoop_CM_2018,Armitage_RMP_2018}. To date, the sharpest observation of topological Weyl semimetals (WSM) has been achieved in a few spatial-inversion symmetry broken materials that yield clear signatures in photo-emission experiments and straightforward identification and classification within {\it ab-initio} calculations \cite{Armitage_RMP_2018}.

\begin{figure}[htp]
\centering
\includegraphics[width=0.5\textwidth]{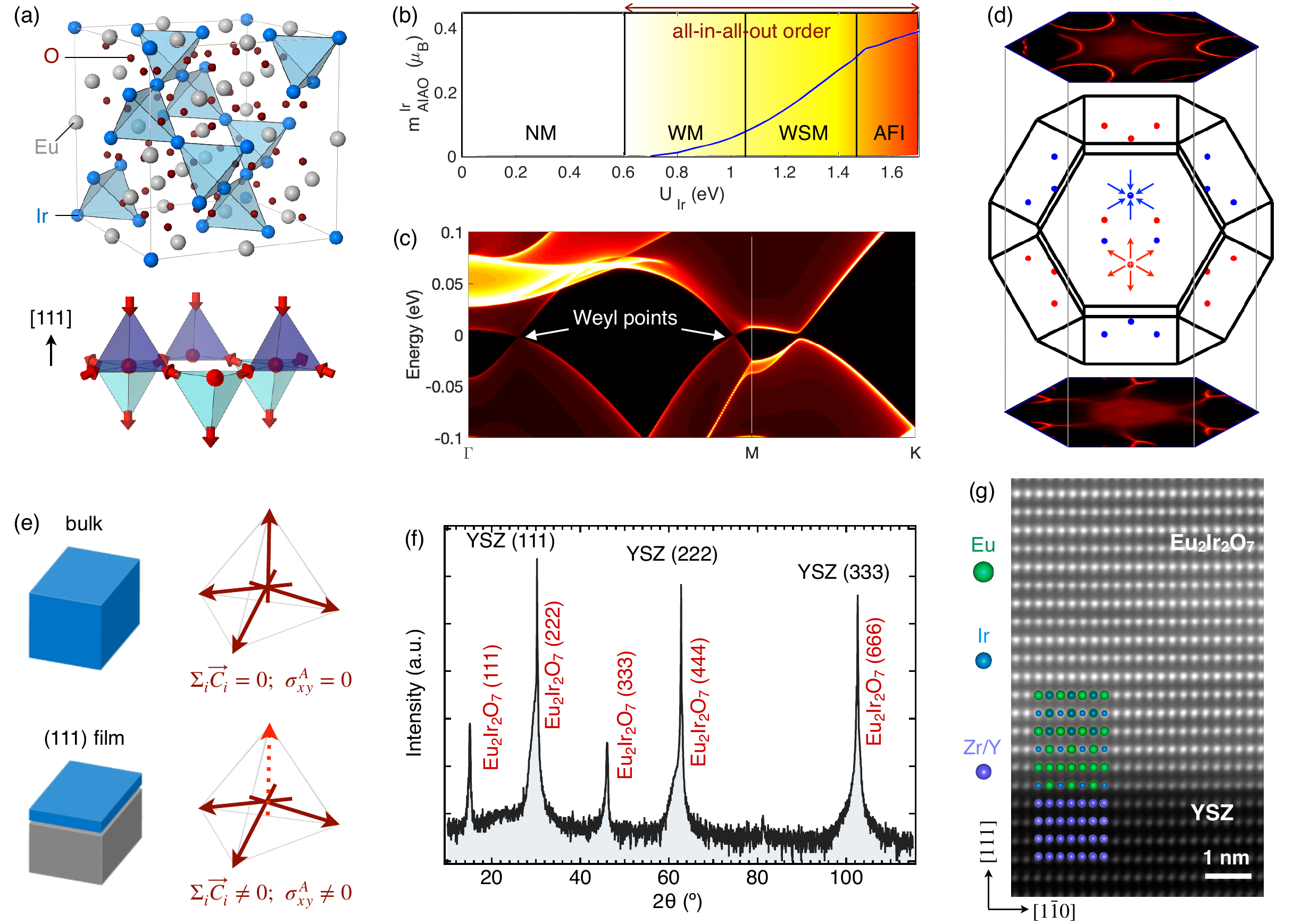}
\caption{\label{structure} 
From bulk Eu$_2$Ir$_2$O$_7$ crystals to high-quality (111) thin films. (a) Unit cell of Eu$_2$Ir$_2$O$_7$ with the Ir corner-sharing tetrahedral framework (top), and the ``all-in-all-out'' antiferromagnetic order (bottom). (b) Schematic phase diagram as a function of Ir local Coulomb interaction $U_{\mathrm{Ir}}$ 
displaying normal metal (NM), Weyl metal (WM), Weyl semimetal (WSM), and antiferromagnetic insulator (AFI). Within DFT+U, Ir all-in-all-out order is sustained when $U_{\mathrm{Ir}} > U_{c1} \approx 0.6$ eV. With increasing $U_{\mathrm{Ir}}$ and magnetic moment $m^{\rm Ir}_{\rm AIAO}$, the WM, WSM and AFI phases are observed with the transition at $U_{c2} \approx 1.05$ and $U_{c3} \approx 1.45$ eV, respectively. (c) Spectral weight of the electronic bands near the Fermi level, along with the Weyl points in the WSM. (d) 24 Weyl points in bulk Brillouin zone (BZ), where the red/blue color denotes the $\chi=\pm 1$ chirality from the Berry curvature flux. The nontrivial topology leads to Fermi arc surface states when the bulk crystal is cleaved. The top (bottom) projections are the spectral plots in the surface BZ cleaved at the kagome (triangle) terminated atomic plane perpendicular to [111]. (e) Emergence of intrinsic anomalous Hall effect in (111) Eu$_2$Ir$_2$O$_7$ thin film as a result of broken cubic symmetry. (f) XRD 2$\theta$-$\omega$ scan of a 40 nm (111) Eu$_2$Ir$_2$O$_7$ thin film grown on YSZ substrate. (g) HAADF-STEM cross-section image of the sample, confirming the expected epitaxial relationship between Eu$_2$Ir$_2$O$_7$ film and YSZ substrate.} 
\end{figure}

In contrast, magnetic WSMs, where the Weyl state emerges due to spontaneous time-reversal symmetry breaking, have remained largely unexplored \cite{Neto_Science_2019}. Pyrochlore iridates with a chemical formula $R_2$Ir$_2$O$_7$ ($R$ = Y or a rare-earth element) are the most theoretically investigated compounds towards this goal, where the entwined band topology, large spin-orbit coupling, and the moderate Coulomb interaction allow for itinerant massless fermions surviving the Mott insulating tendencies \cite{Wan_PRB_2011,WK_PRB_2012, Savary_PRX_2014,Yamaji_PRX_2014,Shinaoka_PRL_2015,Wang_PRB_2017,Berke_NJP_2018}. Particularly, upon lowering the temperature, all $R_2$Ir$_2$O$_7$ except for $R$ = Pr show a transition from the Luttinger-Abrikosov-Beneslavskii (LAB) metallic phase \cite{Moon_PRL_2013} characterized by the quadratic band-touching to a non-metallic ground state \cite{Balents_ARCMP_2014}. Concurrently, the paramagnetic phase transits into an antiferromagnetic (AFM) phase with the all-in-all-out spin configuration stabilized on each cation tetrahedron [Fig.~\ref{structure}(a)]. Such an unusual spin configuration results in two degenerate domain structures, all-in-all-out (AIAO) and all-out-all-in (AOAI), `switchable' by the time-reversal operation \cite{Arima_JPSJ_2013}. Within a single domain of the magnetically ordered state, the time-reversal symmetry is spontaneously broken, and the WSM phase is predicted to emerge by {\it ab-initio} calculations in realistic parameter regions [Fig.~\ref{structure}(b)-(d)] \cite{Wan_PRB_2011,WK_PRB_2012,Shinaoka_PRL_2015,Wang_PRB_2017,Berke_NJP_2018}. On the other hand, despite remarkable experimental efforts, unambiguous demonstrations of the Weyl semimetallic state in this class of materials have remained a challenge \cite{Sushkov_PRB_2015,Ueda_NC_2017,Ueda_NC_2018,LaBarre_JPCM_2020}.

In magnetic WSMs, a non-quantized anomalous Hall conductance (AHC) was specifically predicted, which for one pair of Weyl points is proportional to their distance in the Brillioun zone \cite{Armitage_RMP_2018,Burkov_ARCMP_2018}. 
Thus, the observation of an intrinsic anomalous Hall effect (AHE) accompanied by the all-in-all-out AFM order would serve as one of the direct evidences for the WSM phase in pyrochlore iridates. Surprisingly, while the spin order has been successfully identified in several bulk crystals of pyrochlore iridates, the AHE remained undetected \cite{sagayama_PRB_2013,Donnerer_PRL_2016,Guo_PRB_2016}. To explain the elusive nature of the AHE, a careful examination of the cubic $Fd\bar{3}m$ symmetry of bulk $R_2$Ir$_2$O$_7$ revealed that it enforces a complete cancellation of the Chern vectors from each pair of the Weyl nodes \cite{Nagaosa_PRL_2014} resulting in zero net Hall response. By virtue of the same argument, one can conjecture that breaking the cubic symmetry should lead to incomplete cancellation of the Chern vectors necessary for unveiling a non-zero AHC. Experimentally, this can be realized by either applying a uniaxial strain \cite{Ran_PRB_2011} or by confining the pyrochlore lattice into a slab geometry, along the [111] direction \cite{Nagaosa_PRL_2014,Kim_SciRep_2016}. Both approaches necessitate the (111)-oriented epitaxial thin films as a crucial platform to unwrap the topological features in pyrochlore iridates [Fig.~\ref{structure}(e)] \cite{JC_APLMater_2020}. Here, by using synchrotron-based resonant scattering and absorption techniques, magneto-transport measurements, and first-principles calculations, we report on the discovery of an emergent AHE with a colossal coercive field in the all-in-all-out AFM ordered phase of (111) Eu$_2$Ir$_2$O$_7$ thin films. Our results confirm that these phenomena are stark manifestations of the electronic bands endowed with Weyl crossings in the film geometry with broken cubic symmetry.\\

From the materials standpoint, we select Eu$_2$Ir$_2$O$_7$ primarily for two reasons: (1) The Eu$^{3+}$ ion is non-magnetic, such that the observed properties are exclusively attributed to the Ir$^{4+}$ sublattice \cite{Kim_SA_2020}, and (2) previous studies on bulk Eu$_2$Ir$_2$O$_7$ have shown hints of the WSM phase \cite{Tafti_PRB_2012,Sushkov_PRB_2015,Telang_PRB_2019}. The (111) Eu$_2$Ir$_2$O$_7$ thin films ($\sim$40 nm) were fabricated on (111) yttria-stabilized ZrO$_2$ (YSZ) substrate by pulsed laser deposition using the ``\textit{in-situ} solid phase epitaxy'' method \cite{XL_APL_2020}. High-resolution x-ray diffraction 2$\theta$-$\omega$ scan over a wide range of angles confirms the formation of a pure pyrochlore phase along the (111) orientation [Fig.~\ref{structure}(f)]. A cross-section image from scanning transmission electron microscopy (STEM) further validates the structural quality of the film with the expected epitaxial relationship and atomically sharp film-substrate interface without any buffered layer [Fig.~\ref{structure}(g)].

\begin{figure}[t]
\centering
\includegraphics[width=0.5\textwidth]{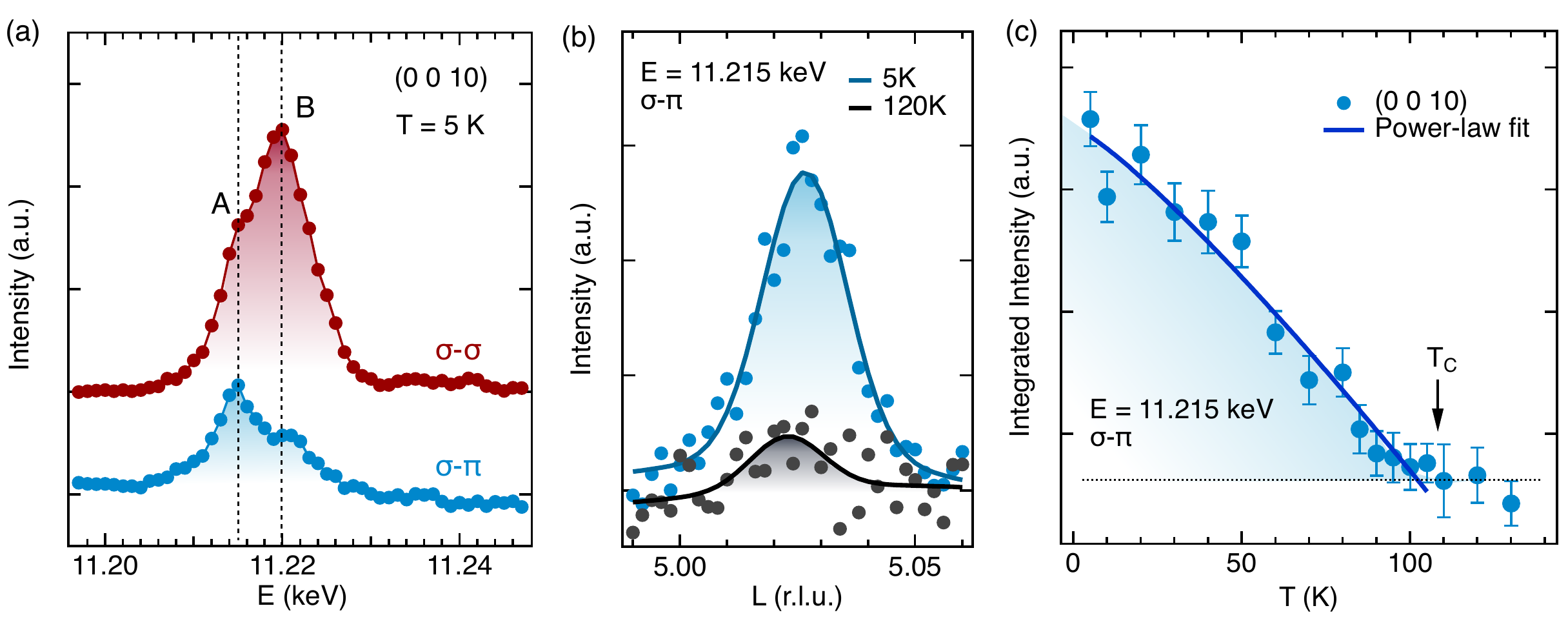}
\caption{\label{RMS} Demonstration of the Ir all-in-all-out ordering. (a) Spectra of the intensity of Eu$_2$Ir$_2$O$_7$ film (0 0 10) reflection near the Ir L$_3$ edge in the $\sigma$-$\sigma$ and the $\sigma$-$\pi$ channels. The enhancement due to resonant magnetic scattering is clearly seen at $\sim$11.215~keV in the $\sigma$-$\pi$ channel. (b) L scans of the (0 0 10) reflection in $\sigma$-$\pi$ channel with a fixed energy at 11.215 keV at high and base temperatures. (c) Temperature dependence of the integrated intensity (over L) of the (0 0 10) reflection. The onset of magnetic phase transition is probed at $T_c$ around 110 K. The blue solid line is a power-law fit of the data below $T_c$, giving rise to the scaling  $\sim |T-T_c|^{1.1}$.}
\end{figure}

Next, we turn to investigate the low-temperature spin structure of the film. It is noteworthy that even for bulk iridate crystals, due to the high absorption cross-section of Ir, neutron scattering is hardly applicable. Alternatively, the magnetic structure can be probed by synchrotron-based x-ray resonant scattering. Recently, the long-range all-in-all-out AFM order on the Ir sublattice has been reported in bulk $R_2$Ir$_2$O$_7$ ($R$ = Sm, Eu, Nd) \cite{sagayama_PRB_2013,Donnerer_PRL_2016,Guo_PRB_2016}. However, such a spin order has never been directly demonstrated in their thin films, and is only presumed from the bulk results. Here, we show the first experimental demonstration of the all-in-all-out order in our (111) Eu$_2$Ir$_2$O$_7$ thin films by x-ray resonant magnetic scattering (XRMS). 

Several specific details of the scattering measurements are due. First, the all-in-all-out spin structure is a {\bf k} = 0 magnetic order, which gives rise to additional (0~0~4n+2) reflections that are structurally forbidden for an ideal pyrochlore lattice with the $Fd\bar{3}m$ space group. However, because of the intrinsic local trigonal distortion of the IrO$_6$ octahedra, there exists an additional charge contribution (known as the anisotropic tensor susceptibility (ATS) scattering) to the (0~0~4n+2) reflections \cite{dmitrienko_AC_2005}. The ATS contribution is dominant in the $\sigma$-$\sigma$ channel, and can be drastically suppressed in the $\sigma$-$\pi$ channel by setting the [011] axis of Eu$_2$Ir$_2$O$_7$ film perpendicular to the scattering plane (see Supplemental Materials).

As seen in Fig.~\ref{RMS}(a), the spectra of the (0 0 10) reflection in both channels exhibit a doublet feature, with the peak position at (A) $E$ = 11.215 and (B) $E$ = 11.22 keV, referring to the resonant excitation from Ir 2$p$ core to $t_{2g}$ and $e_g$ levels, respectively. As the magnetism of Ir$^{4+}$ in pyrochlore iridates predominantly stems from the partially filled $t_{2g}$ levels, magnetic scattering is only resonantly enhanced during the 2$p$-$t_{2g}$ excitation. As a result, the relative intensity of peak A is significantly increased in the $\sigma$-$\pi$ channel, consistent with the results from bulk \cite{sagayama_PRB_2013, Donnerer_PRL_2016}. Next, by fixing the incident energy of x-rays at 11.215 keV, $L$ scans of the (0 0 10) reflection in the $\sigma$-$\pi$ channel reveal the presence of a distinct magnetic Bragg peak at 5 K, which becomes weaker and broader, and eventually disappears into the background near 120 K [Fig.~\ref{RMS}(b)]. These results directly confirm the establishment of the all-in-all-out order in (111) Eu$_2$Ir$_2$O$_7$ thin films. The temperature dependence of the (0 0 10) integrated intensity further reveals the onset of the magnetic transition at $\sim$110 K [Fig.~\ref{RMS}(c)]. Moreover, since the intensity of XRMS in the $\sigma$-$\pi$ channel is proportional to the square of the Ir magnetic moment (denoted as $m^{\rm Ir}_{\rm AIAO}$) \cite{sagayama_PRB_2013}, a fit of the intensity below the transition temperature [solid line in Fig.~\ref{RMS}(c)] yields the power-law behavior  
$(m^{\rm Ir}_{\rm AIAO})^2\sim |T-T_c|^{1.1}$, consistent with a second-order phase transition into the long-range AFM ordered state with a mean-field exponent \cite{ishikawa_PRB_2012}.

\begin{figure}[b]
\centering
\includegraphics[width=0.5\textwidth]{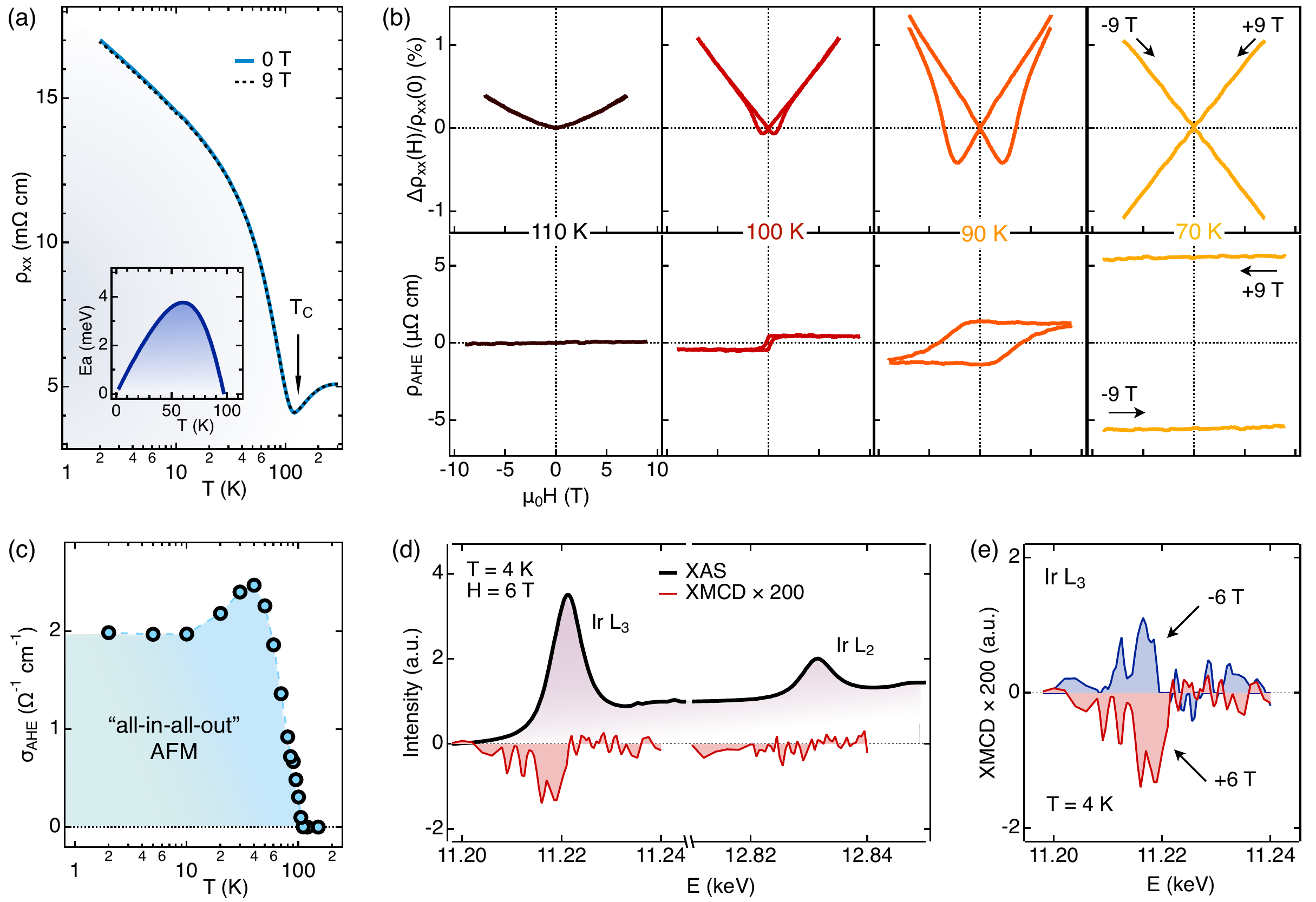}
 \caption{\label{AHE} Intrinsic anomalous Hall effect in (111) Eu$_2$Ir$_2$O$_7$ thin film. (a) Temperature dependence of the longitudinal resistivity under 0 (blue) and 9 T (black) magnetic field, respectively. The onset of transition is found at $T_c$ $\sim$110 K. Inset shows the temperature dependence of the activation gap $E_a$ estimated using the Arrhenius's law. (b) Normalized longitudinal magnetoresistivity and transverse resistivity at a set of temperatures across $T_c$. Note, the linear contribution from the ordinary Hall effect has been subtracted for each transverse curve. (c) Temperature dependence of the anomalous Hall conductivity $\sigma_\textrm{AHE}$. (d) Synchrotron x-ray absorption and magnetic circular dichroism (XMCD) spectra on Ir L$_{3,2}$ edges at 4 K. The magnetic field was applied along the [111] direction, parallel to the surface normal of film. (e) Ir L$_3$ XMCD spectra in +6 T (red) and -6 T (blue), respectively. The sign of XMCD is flipped by direct field sweeping.}
\end{figure}

Now we turn our attention to the key question about the transport properties of our samples. The temperature dependence of longitudinal resistivity $\rho_{xx}$ shows the onset of a transition into the non-metallic phase at $T_c$ $\sim$110 K [Fig.~\ref{AHE}(a)], coincident with the magnetic transition as determined by resonant magnetic scattering. Strikingly, the magnitude of the inverse residual resistivity ratio is rather small, 1/RRR = $\rho$(2K)/$\rho$(300K) $\sim$3.3, and the $\rho$(T) curve yields a rough estimate for a putative Arrhenius activation gap $E_a$ $\leq$ 4 meV [Fig.~\ref{AHE}(a) inset]. These data validate the semi-metallic nature of the (111) Eu$_2$Ir$_2$O$_7$ thin film below $T_c$. In addition, unlike Nd$_2$Ir$_2$O$_7$ with highly metallic domain walls between the AIAO and AOAI magnetic domains \cite{Ueda_PRB_2014,Ma_Science_2015}, the domain walls in Eu$_2$Ir$_2$O$_7$ film contribute no discernible conducting channels, as revealed by the practically identical $\rho_{xx}$(T) behaviors after `training' the film in 0 T (solid blue curve) and 9 T (dotted black curve) magnetic field applied along the (111) direction.

Fig.~\ref{AHE}(b) displays prototypical normalized $\rho_{xx}$ curves as a function of field taken across the transition temperature. While above 110 K the metallic phase possesses a simple parabolic-dependent $\rho_{xx}$(H) shape, a hysteretic butterfly-dependent $\rho_{xx}$(H) takes shape at the onset of the transition, whose magnetic coercivity $H_C$ and saturation $H_S$ fields dramatically increase in a narrow temperature window from 100 to 90 K. At 70 K, due to colossal $H_C$ and $H_S$, only a small portion of the full hysteresis (the `crossing') remains accessible in the measurement. It is important to note, the key part of the protocol to obtain the correct behavior of $\rho_{xx}$(H), is to measure each branch of the hysteresis by thermally quenching the magnetic ordering (see Supplemental Materials). For example, a magneto-transport response observed by Fujita {\it et al.} on a 70 nm (111) Eu$_2$Ir$_2$O$_7$ film was attributed to the exotic ``odd-parity'' magnetoresistance \cite{fujita_SciRep_2015}. Observation of the colossal coercivity corroborates the predicted magnetic {\it octupole} nature of the all-in-all-out order \cite{Arima_JPSJ_2013}, which couples weakly to the external field; as a consequence, the domains are hardly switchable by the direct field sweeping at lower temperatures \cite{Liang_NatPhys_2017,Fujita_APL_2016}. 

Furthermore, following the prediction of AHE in the WSM phase of $R_2$Ir$_2$O$_7$ thin films, a finite signal $\rho_\textrm{AHE}$(H) emerges right below $T_c$. The magnitude of the AHE, $H_C$ and $H_S$ rapidly increase with lowering temperatures [Fig.~\ref{AHE}(b)]. This remarkable feature immediately links the observed AHE to the presence of the long-range all-in-all-out AFM order on the Ir sublattice. The complete set of $\rho_\textrm{AHE}$(H) curves down to 2 K are shown in the Supplemental Materials. Fig.~\ref{AHE}(c) exhibits the deduced anomalous Hall conductivity, $\sigma_\textrm{AHE} = \rho_{yx} / \rho_{xx}^2$, as a function of temperature. As clearly seen, below 10 K, $\sigma_\textrm{AHE}$ levels off at $\sim$2 $\Omega^{-1}$cm$^{-1}$, which corresponds to a net AHC $\sim$0.23 $e^2/h$.

Unlike ferromagnetic metals, where $\sigma_\textrm{AHE}$ is proportional to net magnetization \cite{Nagaosa_RMP_2010}, the appearance of AHE in a completely compensated AFM is highly nontrivial. Therefore, in the discussion about the origin of this emergent AHE in (111) Eu$_2$Ir$_2$O$_7$ thin films, it is critical to rule out the net sample magnetization. For this purpose, we obtained the resonant x-ray absorption spectra (XAS) taken with left- and right-circularly polarized photons on the Ir L$_{3,2}$ edges at 4 K. The magnetization of Ir can be deduced from the difference between those two spectra, the x-ray magnetic circular dichroism (XMCD) [Fig.~\ref{AHE}(d)].
A direct inspection of the XMCD signal shows that even under 6 T magnetic field along the [111] direction, the dichroic signal exhibits a tiny finite intensity at Ir L$_3$ edge and a negligible signal at L$_2$ edge. Remarkably, the sum rules analyses on the spectra yield a net magnetization of only $\sim$0.009(6) $\mu_B$/Ir (Supplemental Materials). 
Such a minute value implies the fully compensated all-in-all-out order, where the Ir spins on each tetrahedron point along the local $\langle111\rangle$ axis leading to the expected perfect moment cancellation and thus zero net magnetization. Crucially, in sharp contrast to the low-temperature AHE behavior, the direction of magnetization is switchable by flipping the orientation of the magnetic field at 4 K, as indicated by the flip of XMCD sign at Ir L$_3$ edge [Fig.~\ref{AHE}(e)]. Overall, these results lend strong support on the unconventional nature of the observed AHE and testify for the nontrivial topology of the electronic bands in momentum space.


\begin{figure}[t]
\centering
\includegraphics[width=0.5\textwidth]{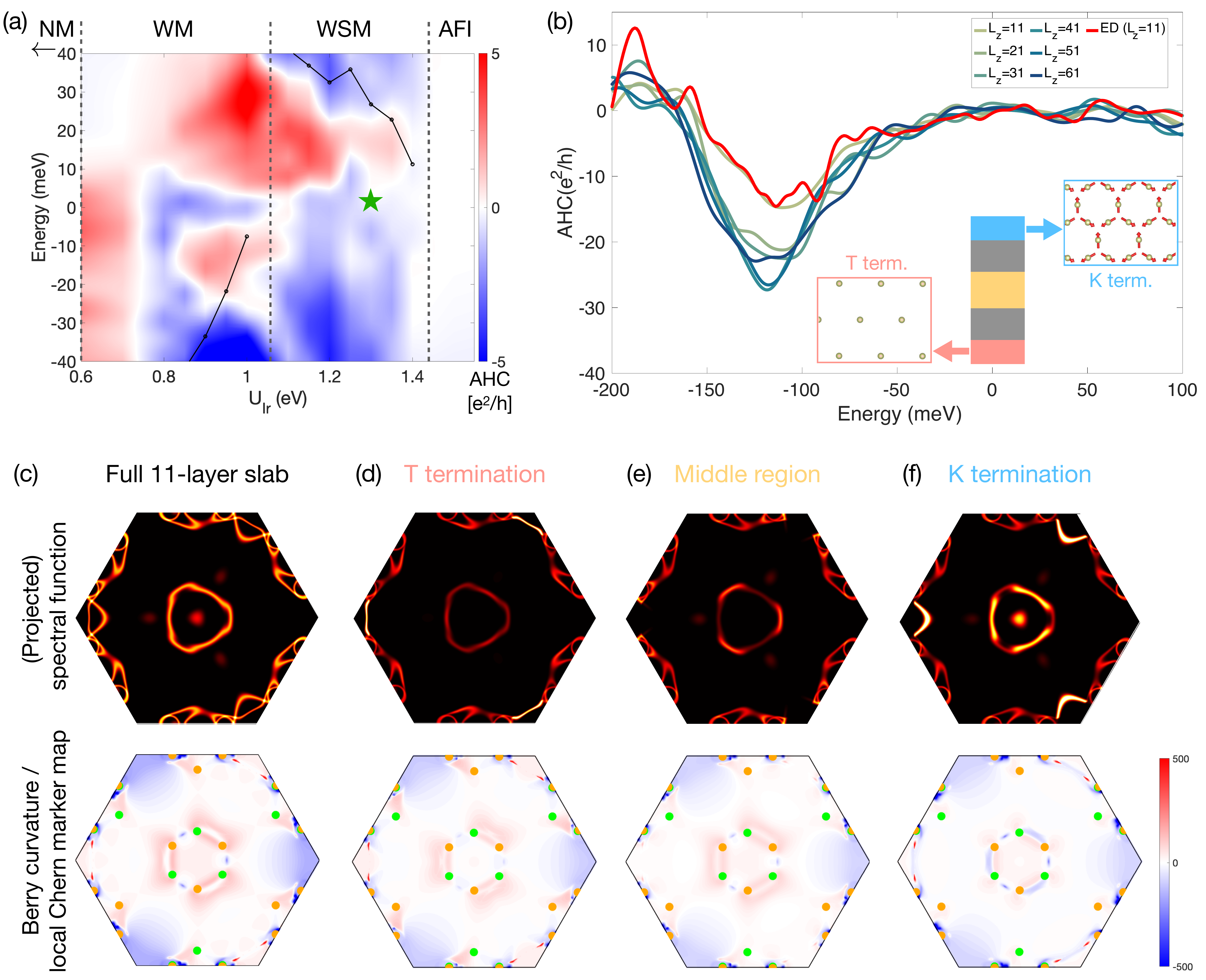}
\caption{\label{AHE_Theory} Electronic structure simulations and the anomalous Hall conductance for (111) Eu$_2$Ir$_2$O$_7$ thin slabs. (a) Anomalous Hall conductance map as functions of U$_{\rm Ir}$ and energy with 11-layer slabs. The black lines indicate the Weyl point energy locations, for both WM and WsM phases. (b) L$_z$ scaling of anomalous Hall conductance for slabs with $U_{\mathrm{Ir}} = 1.3 $ eV based on calculations using the kernel polynomial method. At $L_z=11$, the result from exact diagonalization (red) is also compared. (c) The momentum-dependent spectral function at the Fermi energy ($E = 0$) for an 11-layer slab at $U_{\mathrm{Ir}} = 1.3$ eV, and the associated Berry curvature map. (d-f) Decomposed contributions into triangle (T) / kagome (K) termination regions, and the middle region. In the projected local density of states at the surfaces, the Fermi arc features can be identified. The layer- and momentum-resolved Berry curvature maps are based on local Chern markers~\cite{Chern_marker}.}
\end{figure}

To shed light on the microscopic connection between the observed phenomena and the electronic band topology, we employ first principle calculations of Eu$_2$Ir$_2$O$_7$ in a thin film geometry to directly model the experiment. Within the DFT+U approximation \cite{Wan_PRB_2011}, the bulk electronic phase supports the non-collinear all-in-all-out order with Weyl points in the effective band structure when the onsite Hubbard interaction of Ir $U_{\mathrm{Ir}} > U_{c1}$ [Fig.~\ref{structure}(b)]. As $U_{\mathrm{Ir}}$ increases, the magnitude of the local moment $m^{\rm Ir}_{\rm AIAO}$ also increases. By varying $U_{\mathrm{Ir}}$, the electronic ground state can be in a normal metal (NM), a Weyl metal (WM, with 8 Weyl points concomitant with metallic bands at the Fermi energy), a Weyl semi-metal (WSM, with 24 Weyl points), or a trivial antiferromagnetic insulator (AFI) phase. Importantly, this is qualitatively consistent with results that treat electron correlations beyond DFT+U~\cite{Millis_PRB_2017}. In Fig.~\ref{structure}(d), we show the 24 Weyl points at $U = 1.15$ eV (WSM), which occur at $E_W = 23$ meV. The cubic symmetry, as indicated by the locations of the Weyl points, enforces zero AHC in any direction in a bulk sample \cite{Ran_PRB_2011}.  

The thin film geometry used in the experiments naturally breaks the cubic symmetry by selecting a preferential axis to define the surface plane, and can thus induce a non-zero AHE~\cite{Nagaosa_PRL_2014}. To describe the realistic thin films structure, we constructed a slab model with $L_z$ layers along the $[111]$ direction that is perpendicular to the surface. For the (111)-oriented films, the top (bottom) terminated surfaces are kagome (triangle) Ir atomic planes, respectively, which retains both the proper stoichiometry and critically the vanishing net magnetization under a perfect all-in-all-out AFM order as suggested from experiment (the unbalanced slabs with same terminating lattices and net magnetization are described in the Supplemental Materials). 
Representative results of the calculated AHC for an 11-layer slab are displayed in Fig.~\ref{AHE_Theory}(a), as a function of $U_{\mathrm{Ir}}$ and energy across each phase identified in the bulk simulations. As anticipated, a much more pronounced AHC response is revealed in the WM and WSM phases, whereas in the AFI phase, the AHC is strongly suppressed due to the removal of the Weyl nodes that act as sources and sinks of Berry curvature.  
In addition, it is noteworthy that the Weyl physics in the WM phase can be masked by the enhanced density of states from parasitic metallic bands at the same Fermi energy as the Weyl nodes, which produce a much larger longitudinal conductance (see Supplemental Materials). However, this is at odds with the observed transport results indicating a semi-conducting/semi-metallic behavior below $T_c$. Thus, the ground state of Eu$_2$Ir$_2$O$_7$ is in the WSM regime, as indicated by the star marker. 

Evolution of the simulated AHC in the WSM phase ($U_{\mathrm{Ir}}$ = 1.3 eV) with varying film thickness is shown in Fig.~\ref{AHE_Theory}(b), where $L_z$ is sampled up to 61 layers
that is comparable to the thickness of our experimental films. It is striking that slabs with various $L_z$ almost exhibit a thickness independent behavior, leading to a primarily constant net Hall conductance in the vicinity of the Fermi energy. This produces a Hall conductivity that scales like $\sigma_{xy}\sim 1/L_z$, which will consistently recover the zero response in bulk imposed by the cubic symmetry. 

We further explore the origin of AHC by calculating the momentum dependent spectral function and the Berry curvature through the local Chern marker~\cite{Chern_marker} map. In particular, we compare the total response of the slab in Fig.~\ref{AHE_Theory}(c) with individual contributions [Fig.~\ref{AHE_Theory}(d)-(f)] decomposed into different regions of the slab, as depicted in Fig.\ref{AHE_Theory}(b) inset. 
We observe two distinct contributions to the Hall conductance. First, the major contribution comes from the smooth variations of the local Berry curvature on the surface and in the bulk, which is associated with the quantization of bulk-like states due to the finite slab geometry that form effective quantum well states. 
In addition, by comparing to the surface spectral function, there is a novel minor contribution from the Fermi arc surface states, as revealed by the enhanced surface Berry curvature emanating out of the Weyl node projections on the surface, which is tightly confined in momentum space.\\

In summary, the combined experimental and theoretical results demonstrate the emergence of the long-sought magnetic Weyl semimetal in (111) Eu$_2$Ir$_2$O$_7$ thin films. These findings identify thin films of pyrochlore iridates as a stimulating ground for realizing topologically non-trivial states (e.g. quantum AHE in the quasi-2D limit, axionic insulators, topological magnons), and may open new prospects for the nascent field of topological antiferromagnetic spintronics.\\

The authors deeply acknowledge D. Puggioni, J. M. Rondinelli, A. J. Millis, P. Kissin, R. Averitt, D. Khomskii, G. Fiete, D. Vanderbilt for numerous insightful discussions. X.L. and J.C. acknowledge the support by the Gordon and Betty Moore Foundation EPiQS Initiative through Grant No. GBMF4534, and the Department of Energy under Grant No. DE-SC0012375. Y.F. and J.H.P are partially supported by NSF CAREER Grant No. DMR-1941569. S.F. is supported by a Rutgers Center for Material Theory Distinguished Postdoctoral Fellowship. For the computing, we used the Beowulf cluster at the Department of Physics and Astronomy of Rutgers University. W.G. and W.W. are supported by DOE BES under award DE-SC0018153. This research used resources of the Advanced Photon Source, a U.S. Department of Energy Office of Science User Facility operated by Argonne National Laboratory under Contract No. DE-AC02-06CH11357. A portion of this work was performed at the National High Magnetic Field Laboratory, which is supported by the National Science Foundation Cooperative Agreement No. DMR-1644779 and the state of Florida. The Flatiron Institute is a division of the Simons Foundation.\

\end{document}